\documentstyle{article}
\begin{document}
\newcommand{\nc}{\newcommand}



\nc{\beq}{\begin{equation}}
\nc{\eeq}{\end{equation}}
\nc{\bea}{\begin{eqnarray}}
\nc{\eea}{\end{eqnarray}}
\nc{\artref}[5]{\bibitem{#1} {#2}, {#3} {#4} #5}
\nc{\bookref}[4]{\bibitem{#1} {#2}, {#3}, #4}
\nc{\fn}{\footnote}


\nc{\gvt}{\vartheta}
\nc{\gt}{\theta}
\nc{\gct}{\Theta}
\nc{\gvp}{\varphi}
\nc{\gp}{\phi}
\nc{\gcp}{\Phi}


\nc{\stw}{\sin (\gvt _w)}
\nc{\ctw}{\cos (\gvt _w)}
\nc{\hisym}{SU(2)_L \otimes SU(2)_R}
\nc{\cusym}{SU(2)_C}
\nc{\sul}{SU(2)_L}
\nc{\gasym}{\sul\otimes U(1)_Y}
\nc{\fpp}{{H_5^{++}}   }
\nc{\fp }{{H_5^+}      }
\nc{\fne}{{H_5^o}      }
\nc{\tp }{{H_3^+}      }
\nc{\tne}{{H_3^o}      }
\nc{\sne}{{H_1^o}      }
\nc{\snp}{{H_{1^\prime}^o}  }


\begin{titlepage}
\thispagestyle{empty}

\begin{tabular}{l}
\hspace{11.3cm}  NEIP93-002\\
\hspace{11.3cm}  ETH-TH/93-8\\
\hspace{11.3cm}  3. March 1993\\
\end{tabular}\\

\vspace{1.0 truecm}

\begin{center}
{\bf \Large{
Gauge boson masses dominantly generated\\
by Higgs-triplet contributions ?}
}\end{center}

\vspace{1.5 truecm}

\centerline{\large\bf\sc Peter Bamert}
\bigskip\bigskip
\centerline{\it Institut de Physique}
\centerline{\it Universit\' e de Neuch\^ atel}
\centerline{\it CH-2000 Neuch\^ atel, Switzerland}
\bigskip\bigskip
\centerline{and }
\bigskip\bigskip\bigskip
\centerline{\large\bf\sc Zoltan Kunszt}
\bigskip\bigskip
\centerline{\it Institute of Theoretical Physics, ETH,}
\centerline{\it CH-8093 Z\" urich, Switzerland}

\vspace{2.5 truecm}

\begin{abstract}
\bigskip
\noindent
We discuss a model in which the Standard Model (SM) Higgs
sector has been extended by  additional real and complex
triplets. The $\rho\approx 1$ constraint is satisfied by
restricting the potential to have an enlarged $\hisym $
global symmetry. This is fine tuning, which leads to a
decreased predictability in next-to-leading order.
In this model, however,
the triplet vacuum expectation values
may give the dominating contribution to the gauge
boson masses.
Using a renormalization group argument we constrain
this region of the parameter space.
Another interesting feature of this model is that
one of the neutral scalars
doesn't couple to the fermion sector at tree level and
therefore could have a
relatively large branching ratio to $2\gamma $'s.
It is coupled, however, to the $Z$-boson and
therefore it could
be produced at LEP via the standard $e^+e^-\rightarrow
Z\phi$ mechanism
with rates
comparable to the ones  of the Standard Model.
\end{abstract}

%
%

\end{titlepage}



The Standard Model (SM) of electroweak and strong interactions
successfully describes all elementary particle physics
phenomena.
Recently, it has been successfully  confronted with
the precision measurements at LEP \cite{ROLANDI}.
The Higgs sector, and so the mechanism of the
electroweak symmetry breaking, however, is
weakly constrained by the data. The most important constraint
is given by the measured value of the rho parameter:
\bea \rho & = & {{M_W}^2\over {M_Z}^2 \cos^2(\gvt _w)} \\
{\rho }_{exp} & = & 1.003 \pm
0.004\footnotemark\;\;\;\;\cite{RPP}\; . \nonumber \eea
Models\footnotetext{Fit of experimental data with
$M_{top}=100\; GeV$.} involving only $\sul$ doublets and
singlets, satisfy in a natural way the
$\rho\approx 1$ condition. This is due to the fact that
in the  SM the $\gasym $ gauge symmetry forces the
scalar potential to exhibit a $\hisym $ global symmetry.
Breaking the gauge symmetry the Higgs potential still exhibits
an $\cusym $ ({\it custodial}) symmetry which
ensures $\rho =1$ at tree level. Because of this, $\hisym $
violating radiative corrections to the Higgs potential are
finite and small. This also means that the rho parameter
is close to one naturally, without fine tuning parameters.

If we want to allow also for Higgs triplets or even
higher multiplets, the value of the measured rho parameter
becomes a strong constraint. One obvious possibility
is that we demand the triplet vacuum expectation values
(VEV's) to be small. This requires fine tuning and an effective
decoupling of the triplets from gauge bosons and fermions.
Alternatively, we
could require that the scalar potential, even with these
higher multiplets, will remain invariant with respect to the
enlarged  $\hisym $ global symmetry
\cite{GM,CG}. For triplets, however,  the $\hisym $
symmetry of the Higgs potential will not be obtained automatically
as a result of the gauge symmetry. It can only be achieved by
fine tuning the parameters of the Higgs potential.
Therefore,
the $\rho$-parameter in such models  will  suffer
 from a fine tuning problem and
will depend "directly" on the fine-tuned parameters
 of the scalar potential.
In the SM model the $\rho$ parameter is protected from
this problem by the custodial symmetry.

With  fine tuned $\hisym$ symmetry, however,   the higher
dimensional multiplets  may generate a
major part of the W and Z masses.  Furthermore, in these models
the doublet Yukawa couplings could be strongly enhanced
with respect to the SM, indicating an interesting
phenomenology. In this note we shall discuss the phenomenological
viability of this scenario.


 The most straightforward extension of the SM Higgs
sector, realizing the above situation, has been presented by
Georgy and Machacek \cite{GM} who added a complex and a real
triplet to the SM doublet. With this content of the scalar
sector it is indeed possible to restrict the potential to a
$\hisym $ symmetric version, as has been investigated in some
detail by Chanowitz and Golden \cite{CG}. More recent papers
\cite{HH,GVWMOD,GVWREN} have taken a further look at the
renormalization problems \cite{GVWREN} and the phenomenology
\cite{HH,GVWMOD} of this model.

  From the tree level point of view the Higgs sector presents
itself as follows \cite{GM}
\beq \label{sulfields} \Phi = \left( \begin{array}{cc}
                   {\phi^o}^* & \phi^+ \\
                   \phi^- & \phi^o \end{array} \right)\;\; ,\;\;
     X = \left( \begin{array}{ccc}
                   \chi^o & \xi^+ & \chi^{++} \\
                   \chi^- & \xi^o & \chi^+ \\
                   \chi^{--} & \xi^- & {\chi^o}^* \end{array}\right)
                   \eeq
denoting a $({1\over 2},{1\over 2})$ and a $(1,1)$ multiplet
of $\hisym $, with hypercharge assignments of $Y=1$, $0$ and
$2$ for the doublet, the real and the complex triplet
respectively and with the following phase conventions: $\phi^-
\! =\! -{\phi^+}^*$, $\chi^- \! =\! -{\chi^+}^*$, $\xi^-\! =\!
-{\xi^+}^*$, $\chi^{--}\! =\! {\chi^{++}}^*$ and $\xi^o =
{\xi^o}^*$. They transform under $\hisym $ global
transformations as $\Phi\rightarrow U_L\Phi {U_R}^\dagger $
($\Phi\leftrightarrow X$), with $U_L$,($U_R$) being the
transformation matrices in the appropriate representation. The
restricted Higgs potential takes the form \cite{CG}
\bea \label{potential} V(\Phi ,X)\!\!\! &=&\!\!\!
\lambda_1(Tr[\Phi^\dagger\Phi ]-a^2)^2 \nonumber \\
                &+&\!\!\! \lambda_2(Tr[X^\dagger X]-3b^2)^2
\nonumber \\
                &+&\!\!\! \lambda_3(Tr[\Phi^\dagger\Phi]-a^2+
                Tr[X^\dagger X]-3b^2)^2 \\
                &+&\!\!\! \lambda_4(Tr[\Phi^\dagger\Phi
]Tr[X^\dagger X]- 2Tr[\Phi^\dagger T^i\Phi
T^j]Tr[X^\dagger
                    T^iXT^j]) \nonumber \\
                &+&\!\!\!\lambda_5(3Tr(X^\dagger XX^\dagger
X]\! -\!
                   Tr[X^\dagger X]^2)\; , \nonumber \eea
where the sum is taken over $i$ and $j$. The vacuum states
$|\Phi >_o = {a\over\sqrt 2}${\bf 1} and $|X>_o = b${\bf 1}
are defined by $V(|\Phi >_o,|X>_o)=0$ together with the
positivity conditions \cite{CG}
\bea \label{pos} \lambda_1+\lambda_2+2\lambda_3 &\geq & 0
\nonumber \\
\lambda_1\lambda_2 +\lambda_1\lambda_3 +\lambda_2\lambda_3
&\geq & 0 \nonumber \\
\lambda_4 &\geq & 0 \\
\lambda_5 &\geq & 0 \; . \nonumber \eea
The two VEV's $a$ and $b$ are related through the $W$-mass
\beq {M_W}^2 = {M_Z}^2\cos^2\gvt_w = {g^2\over
4}(a^2+8b^2) = {g^2\over 4}v^2 \eeq
with $g$ being the $\sul $ coupling constant and $v$ ($\sim
250\; GeV$) the VEV of the SM Higgs. This allows the
definition of a mixing angle $\gt _H$ denoted by its cosine $c_H$,
sine $s_H$ (and tangent $t_H\equiv\tan (\gt _H)$)
\beq\label{ch} c_H={a\over v}\; ,\;\;\; s_H = {2{\sqrt
2}b\over v} \; . \eeq
Within the Higgs sector the gauged $\gasym $ can be regarded
as a subgroup of $\hisym $ with $T^3_R$ playing the role of
the hypercharge. A diagonal subgroup of $\hisym $ survives the
spontaneous symmetry breakdown. The physical Higgs bosons will
form multiplets, which are degenerate in mass, under this {\it
custodial} $\cusym $ global symmetry.
Expressed in terms of the fields defined in
eq(\ref{sulfields}) they are\large
\beq\label{cusfields}
\begin{array}{ccc}
     \fpp &=& \chi^{++}\hfill \vspace{0.2truecm} \\
     \fp  &=& {1\over\sqrt 2}(\chi^+-\xi^-)\hfill\vspace{0.2truecm}\\
     \fne &=& {1\over\sqrt6}(\chi^o+{\chi^o}^*-2\xi^o)\hfill
\vspace{0.2truecm}\\
     \snp &=& {(\chi^o+{\chi^o}^*+\xi^o)\over\sqrt 3}
\hfill\end{array}
\begin{array}{ccc}
     \tp  &=& c_H{(\chi^++\xi^+)\over\sqrt2}-s_H\phi^+
\hfill\vspace{0.2truecm}\\
     \tne &=&
c_H{(\chi^o-{\chi^o}^*)\over\sqrt2}+s_H{(\phi^o-{\phi^o}^*)
\over\sqrt 2}\hfill\vspace{0.2truecm}\\
     \sne &=& {(\phi^o+{\phi^o}^*)\over\sqrt2}\hfill
\vspace{0.2truecm}\\
     \end{array}\eeq\normalsize
with $H_5$, $H_3$, $H_1$ and $H_{1^\prime}$ denoting a $\cusym
$ fiveplet, triplet and two singlets respectively. The
Goldstone triplet is orthogonal to the $H_3$-plet
\bea \label{goldstone} G_3^+ \!\! &=&\!\!
s_H{(\chi^++\xi^+)\over\sqrt 2}+c_H\phi^+\nonumber\\
G_3^o \!\!&=&\!\! s_H{(\chi^0-{\chi^o}^*)\over\sqrt
2}-c_H{(\phi^o-{\phi^o}^*)\over\sqrt 2} \; .\eea
In terms of those fields one sees that the quantity $c_H$
defined in eq(\ref{ch}) indeed denotes the cosine of a $\sul $
multiplet rotation to $\cusym $ triplets.
The masses are\fn{Note that the positivity of the squared masses
is guaranteed by eq(\ref{pos}).} \cite{CG}
\bea\label{masses}
M_{H_5}^2 &=& 3v^2(c_H^2\lambda_4 +s_H^2\lambda_5) \nonumber
\\
M_{H_3}^2 &=& v^2\lambda_4 \\
M_{H_1,H_{1^\prime}}^2\!\!\!\! &=& \!\!\! 8v^2\left( \!\!
\begin{array}{cc}
                c_H^2(\lambda_1+\lambda_3) & {\sqrt{3\over
8}}c_Hs_H\lambda_3 \\
  {\sqrt{3\over 8}}c_Hs_H\lambda_3 & {3\over
8}s_H^2(\lambda_2+\lambda_3) \end{array} \!\!\right) \; .\nonumber
\eea
It's convenient to rotate the two $\cusym $ singlets to their
mass eigenstates $\tilde{\sne }$ and $\tilde{\snp }$. The
rotation angle shall be denoted by its cosine $c_S$ and sine
$s_S$. $\tilde{\sne }$,($\tilde{\snp }$) defines the mass
eigenstates with the larger $\cusym $ doublet ($\sne
$),(triplet ($\snp $)) portion\fn{Note that by means of this
definition $s_S$ takes values between $-{1\over\sqrt 2}$ and
$+{1\over\sqrt 2}$.}.

Here we consider only Dirac type Yukawa couplings\fn{Majorana
type Yukawa couplings induced by the $\sul $ triplets can be
introduced but are unlikely to be phenomenologically important
unless $t_H\ll 1$ \cite{GVWMOD}.}. Of course, only those fields
of eq(\ref{cusfields}) which have a nonzero doublet admixture
will couple (at tree level) to the fermion sector. Due to
$a\leq v$ the Yukawa couplings are enhanced compared with the
SM. They are\fn{For simplicity we have assumed no singlet
mixing ($c_S=1$) in the following expressions.} (omitting an
overall factor $ig$ ).
\bea\label{yukawa} \sne f\bar{f} \!\! &:& -{M_f\over
2M_W}{1\over c_H} \nonumber\\
\tne f\bar{f}\!\! &:& \pm {M_f\over 2M_W}t_H\gamma_5 \nonumber
\\
\tp\bar{\nu }e \!\! &:& {M_f\over
2M_W}t_H{(1+\gamma_5)\over\sqrt 2} \\
\tp\bar{p}n \!\! &:& {t_HV_{pn}\over 2M_W}\left[
M_n{(1+\gamma_5)\over\sqrt 2} - M_p{(1-\gamma_5)\over\sqrt
2}\right]\; , \nonumber \eea
where the $\pm $ in the second term refers to up, down type
fermions, $p$ and $n$ denote up and down type quarks and, $\nu
$ and $e$ stand for neutrino type and electron type leptons
respectively. $V_{pn}$ is the CKM mixing matrix.
In the following let $V$ denote either a $W$ or a $Z$ boson.
The VVH type vertices of this model are (omitting an overall
factor $igg^{\mu\nu }$)\cite{GM} :\large
\beq \label{VVH}
\begin{array}{ccc}
W^+W^+H_5^{--} &:& {\sqrt 2}M_Ws_H\hfill \vspace{0.2truecm} \\
W^+W^-\fne &:& -{M_W\over\sqrt 3}s_H\hfill \vspace{0.2truecm} \\
W^+W^-\snp &:& {2{\sqrt 2}\over\sqrt 3}M_Ws_H\hfill
\vspace{0.2truecm}\\
W^+W^-\sne &:& M_Wc_H \hfill \\ \end{array}
\;\;\;\begin{array}{ccc}
ZZ\fne &:& {2\over\sqrt 3}{M_W\over c_W^2}s_H\hfill
\vspace{0.2truecm}\\
ZZ\snp &:& 2{\sqrt{2\over 3}}{M_W\over c_W^2}s_H\hfill
\vspace{0.2truecm}\\
ZZ\sne &:& {M_W\over c_W^2}c_H  \hfill\vspace{0.2truecm}\\
W^-Z\fp &:& -{M_W\over c_W}s_H \hfill .\end{array}
\eeq\normalsize
Again the $\sne\snp $ mixing has been taken to be zero. The
absence of $\tne VV$ couplings is due to $\tne $ being CP-odd.

The seven parameters of the restricted potential
(eq(\ref{potential})) can be reexpressed in terms of the
$W$-mass, the 4 Higgs masses, the singlet mixing ($c_S$) and
$t_H$ ($={s_H\over c_H}$).

Let's first have a brief look at the limit "$t_H\!\ll\!
1$" ($b\!\ll\! a\!\sim\! v$). In this limit the mixing
angle of the two singlets approaches zero: $c_S^2 = 1 -
{3\over 8}{\left( {\lambda_3\over
(\lambda_1+\lambda_3)}\right)}^2t_H^2 +{\cal O}(t_H^4)$. The $\sul $
doublet VEV generates the gauge boson masses and the Goldstone
bosons are the same as in the SM (eq(\ref{goldstone})). The
field $\sne $ has SM $Hff$ and $HVV$ and suppressed $HHV$
couplings, whereas all other Higgs bosons have suppressed
couplings to the fermion sector (eq(\ref{yukawa})) and to the
$HVV$ sector.
The whole situation therefore resembles very much to the SM
with $\sne $ playing the role of the SM Higgs boson. The main
differences come from the non vanishing $HHV$ type
couplings\fn{For a list of the $HHV$ type couplings see e.g.
\cite{GVWMOD}.}, which can serve to bound the masses of the
other (not $\sne $) Higgs fields from below, and the fact that
the other singlet ($\snp $) is likely to be very light:
$M_{H_{1^\prime }}^2 \approx
3v^2s_H^2{(\lambda_1\lambda_2+\lambda_2\lambda_3+\lambda_1
\lambda_3)\over(\lambda_1+\lambda_3)}$.

The  opposite limit "$t_H\!\!\gg\!\! 1$" ($a\!\!\ll\!\! b$) is
far more interesting, because here the $\sul $ triplet fields
($X$) are the source of the SSB, generating the $W$ and $Z$
masses. Again there is almost no singlet mixing: ${c_S^2 = 1
-{8\over 3}{\left( {\lambda_3\over
\lambda_2+\lambda_3}\right)}^2t_H^{-2}+{\cal O}(t_H^{-4})}$.
Here we have the remarkable situation that $\cusym $
multiplets either couple to the fermion sector ($H_1$,$H_3$),
those couplings being strongly enhanced with respect to the
SM, or to the $HVV$ sector ($H_{1^\prime }$,$H_5$), with
couplings being roughly of the same order of magnitude as the
SM ones, but not both. Again there is a field which is likely
to be lighter than the others: ${M_{H_1}^2 \approx
4c_H^2{(\lambda_1\lambda_2 +\lambda_2\lambda_3
+\lambda_1\lambda_3)\over (\lambda_2 +\lambda_3 )}}$.

For medium values of $t_H$ ($\sim {\cal O}(1)$) there can be
significant $\sne\snp $ mixing allowing the mass eigenstates
$\tilde{\sne }$ and $\tilde{\snp }$ to have SM like couplings
in both the fermion and the $HVV$ sector. As in the previous
case $H_3$ couples to the fermion sector while $H_5$ has $HVV$
couplings, but not the other way round.

The restricted potential (eq(\ref{potential})) emerges after
the additional parameters of the general $\gasym $ invariant
potential\fn{For the explicit form of this potential see e.g.
\cite{GVWREN}.} have been fine tuned to be zero\fn{Note that
one cannot simply impose that those terms don't exist, because
they are automatically introduced by means of counterterms
after the renormalization procedure.} on a certain mass scale
(e.g. on the $M_{H_5}$ shell). However, due to the running of
those parameters this {\it fine-tuned custodial symmetry} won't
be exact anymore at another mass scale (e.g. the triplet mass
shell $M_{H_3}$). It is therefore reasonable to understand the
term {\it fine-tuned custodial symmetry} as an approximate
symmetry, which is valid only at tree level. Fine tuning in our
case means that the model looses its predictive power at the
next-to-leading order level\fn{Note that custodial $\cusym $
symmetry is sufficient but not necessary for $\rho =1$. One
can, however, think of this model as an effective theory of a
more fundamental structure from which this tuned property
might emerge on some dynamical reason.}.

At this point one might be tempted to supersymmetrize the model. In
fact, supersymmetry provides a solution to those fine tuning problems
since it ensures that the parameters of the superpotential, from
which the scalar potential derives, do not receive radiative
corrections. This is due to cancellations among various
diagrams ({\it Non Renormalization Theorems}). After
supersymmetry has been softly broken those parameters will
 only receive logarithmic corrections. An extension of
the supersymmetric SM containing one (necessarily complex) triplet
with zero hypercharge $\xi $ has been recently studied
by Espinosa and Quir\' os
\cite{QUIROS}.  In this model, however, the triplet VEV has to be
kept small to fulfill the constraint $\rho\approx 1$.
If we want to have large triplet VEV's,  then  at least one
chiral triplet with nonzero hypercharge has to be added
as becomes clear from the
tree level relation
\beq \label{rhoformula} \rho = {\sum\limits_{\{\Phi\} }^{} {|<\Phi
>_o|}^2(T_L(T_L+1)-(Y/2)^2)\over \sum\limits_{\{\Phi\} }^{} 2{|<\Phi
>_o|}^2(Y/2)^2} \eeq
where the sum goes over all (complex) multiplets, $T_L$ denotes
the largest eigenvalue of the $\sul $ generator $T_3$ in the
appropriate representation and $Y$ is the hypercharge.
Adding e.g. a ${\chi : (3,-2)}$ to ${\xi : (3,0)}$, ${\bar{H} : (2,1)}$
and ${H : (2,-1)}$ [$(a,b)\;\sim\;\gasym $] would require
${|<\xi >_o|={|<\chi >_o|\over\sqrt{2}}}$ for $\rho = 1$. But
since $\chi $ misses its "counterpart" $(3,2)$ the Higgsinos remain
massless and give rise to a chiral anomaly. One then has to introduce
a third chiral triplet ${\bar{\chi } : (3,2)}$. The conditions for
$\rho =1$ are now ${|<\chi >_o|^2 + |<\bar\chi >_o|^2 = 2 |<\xi
>_o|^2}$, resembling, in a special case, to the non supersymmetric
model: ${<\chi >_o}={<\bar{\chi }>_o}={<\xi >_o = b}$.
In this model the scalar Higgs sector has 23 physical
degrees of freedom corresponding
to 2 doubly charged, 5 singly charged and 9 neutral scalars. The most
general gauge invariant superpotential has 8 parameters
(corresponding to the singlet combinations : $H\bar{H}$,
$\chi\bar{\chi }$, $\xi\xi$,
$H\xi\bar{H}$, $\bar{H}\chi\bar{H}$, $H\bar{\chi } H$, $\xi\xi\xi$,
$\chi\xi\bar{\chi } $). The soft breaking terms add another 13
parameters, which are, however, not independent if one assumes that
soft breaking is produced by spontaneous breakdown of local
supersymmetry. In this case the 13 parameters could, in principle, be
computed by renormalizing the 4 independent parameters at the scale
of supergravity breaking down to a low scale. In the end this leaves
us with 12 parameters and 18 masses (including W and Z) indicating
the existence of mass relations. Still there is quite a number of
independent parameters. It is therefore clear that, although
in the supersymmetrized model the
parameters of the
scalar potential can be restricted such that the theory has a vacuum
which satisfies $\rho = 1$ at tree level, with stabilized
fine tuning, such a  model has difficulties with
predictability due to the larger multiplet content and large
number of parameters.

Resuming the discussion of the non supersymmetric
model we see, that the Higgs sector, as presented
before, undergoes minor changes in terms of
an approximate custodial symmetry.
Note that the quantity $t_H$ is only defined in the tree level
approximation, since  the VEV's of $\chi $ and $\xi $ are not
identical in general. Also there will now be mixings
between $\cusym $ representations of different
dimensionalities, leading to a breaking of the mass degeneracy
of the $\cusym $ multiplets. This means that $\rho\neq 1$ in
general, with $(\rho -1)$ being small, of the order of 1-loop
corrections, but not computable, since in this order
additional new parameters of the theory also contribute.
Similarly, new parameters contribute also to
mixing effects breaking the custodial $\cusym $.

In fact, at 1-loop level, all mixings, that respect charge and
CP conservation, among Higgs and gauge boson fields are
allowed now\fn{$\tne $, consisting of the same imaginary field
parts as $G_3^o$, has the same CP assignment (-1), whereas the
other neutral scalars $\tilde{\sne }$, $\tilde{\snp }$ and
$\fne $ have CP (+1). It can be shown directly that the
various 1-loop corrections to e.g. $\tne\fne $ mixing cancel
each other \cite{GVWREN}.}. The fields $\fp $ and $\fne $ for
instance couple now to the fermion sector not only by means of
triangle diagrams, but also by $\fp\tp $ and $\fne\sne $
mixing terms. Decays of those fields to fermion pairs are
therefore likely to be dominant below the $W$-threshold.
Unfortunately the corresponding branching ratios cannot be
computed due to the loss of next-to-leading order
predictability.

When trying to constrain the Higgs sector of this model it
seems reasonable first to concentrate on $t_H$, since it
distinguishes between uninteresting ($t_H\!\ll\! 1$) and
interesting ($t_H\!\gg\! 1$) regions of the parameter space.
Strict upper bounds on $t_H$ can only be got with the help of
unitarity arguments, because one can always render the
enhanced Yukawa couplings phenomenologically negligible by
demanding that the corresponding Higgs masses are big enough.
Tree level partial wave unitarity constraints from $\sne t\bar{t}$
and $\sne\sne $ scattering combined with the lower bound on
$M_{H_1}$ ($\geq \sim 5\; GeV$ \cite{RPP}) coming from the
$\Upsilon $ decay gives roughly $30$ as an upper bound on
$t_H$ \cite{DIPL}. Unitarity limits coming from longitudinal
$W_LW_L$ and $W_LZ_L$ scattering give upper bounds on $\snp $
and $\fne $ masses ($t_H\!\gg\! 1)$ that are similar to the SM
bounds, roughly $M_{1^\prime },M_5\leq 1\; TeV$ \cite{GVWMOD}.
Some ad hoc arguments show what the general unitarity limit on
$t_H$ might look like: Assuming for instance that the
$\lambda_i$ parameters are about similar in magnitude would
yield $M_3\leq 1\; TeV$. Combining this with constraints
coming from $B\bar{B}$ mixing \cite{BBAR1} (see below) results
in limits like $t_H\leq\;\sim 10$.

The strongest bound on $t_H$ derives from a renormalization
group argument, based on two simple assumptions:
"perturbative unification" and "desert". First one wants
perturbation theory to be valid up to a high
scale (e.g. an unification scale $M_U$), furthermore
one excludes new physics in between $M_W$ and $M_U$, so that
the renormalization group equations (RGE) evolve all the way
up to $M_U$ in an effective $SU(3)_C\otimes SU(2)_L\otimes
U(1)_Y$ theory.
The point is that the RGE of the doublet Yukawa couplings have
an infrared fixed point \cite{YUKREN} and that those couplings
inevitably blow up at some high scale if their value is close
to or higher than this fixed point (at the $W$-scale). This
upper bound on the Yukawa couplings gives upper bounds on the
sum of the squared masses of quarks or leptons \cite{YUKREN}. For
the case of three families this results in an upper bound on
the $top$ mass \cite{YUKRENVAL} as
\beq\label{topbound} M_{top}\leq 245\; GeV \; .\eeq
which translates in our case to $M_{top}\leq 245\cdot c_H$ (GeV) or
\beq t_H\leq\sqrt{{\left({245\; GeV\over M_{top}}\right)
}^2-1} \; .\eeq
With given lower bounds on the $top$ mass one gets the
following upper bounds on $t_H$
\bea t_H\leq 2.5 &&{\hbox{\rm for }} M_{top}> 92\; GeV\; ,
\nonumber \\
     \label{th}t_H\leq 4.4 &&{\hbox{\rm for }} M_{top}> 54\;
GeV \; .\eea
The first bound \cite{TOPH} is valid under the assumption that
the
decay $t\rightarrow H^+b$ and other non-SM decays do not
occur. It therefore requires that $M_{H_5}$ and $M_{H_3}$ both
are bigger than $92\; GeV$.
$M_{H_5}$ or $M_{H_3}$ below $92\; GeV$ asks for a decay mode
independent lower bound on the top mass. Such a bound can be
derived from the $W$ width ($M_{top}> 54\; GeV$) \cite{TOPW}.
A low mass charged Higgs boson, therefore, strongly weakens the
unitarity bound on $t_H$. The result given by eq.(\ref{th})
is rather significant: it tells us that if we demand to have
a Higgs sector with Higgs triplets such that the triplet
contribution to the $W$-mass is as large as possible,
the assumptions of "perturbative unification" and "desert"
then restrict the triplet contribution to less than
78\% or 63\%, respectively.

The decay $Z^o\rightarrow H^+H^-$ provides a $t_H$ independent
lower bound on the five- and three-plet masses $M_{H_5}$ and
$M_{H_3}$ \cite{DAVIER} as
\beq \label{hibound} M_{H_5},\; M_{H_3} \geq 41.7\; GeV\; .\eeq
One might believe that, in addition to this high energy bound,
there could also be a low energy bound coming from the mixing
of the $B$ and $\bar{B}$ mesons, since the field $\tp $ can give
a dominant contribution, for high $t_H$, to the flavor
changing neutral currents responsible for this mixing. This
bound depends strongly on $t_H$ (i.e. the Yukawa couplings)
and on $M_{top}$, in the sense that a low $M_{top}$ weakens
the constraint as does a low $t_H$. It turns out that the
unitarity bound on $t_H$ is too strong for the $B\bar{B}$ bound
to take effect. The 3-plet mass $M_{H_3}$ is not constrained
by  $B\bar{B}$ mixing. Results for certain two Higgs doublet
models \cite{BBAR1,BBAR2} can be directly adapted to our case.
Fortunately there is another low energy bound, that does constrain
the triplet mass. It is coming from the bottom decay $b\rightarrow
s\gamma $, which is mediated by penguin diagrams involving the $top$,
$\tp $ and $W$ fields. Again this constraint is weakened for low
$M_{top}$ and low $t_H$. The results obtained in two-Higgs-doublet
models \cite{BSY} apply to our case directly.
The combined constraints on $M_{H_3}$ are shown in
figures(1).

The fiveplet mass $M_{H_5}$ is, in addition to
eq(\ref{hibound}), indirectly constrained by the lower bounds
on $M_{H_3}$. Combining eqs(\ref{pos},\ref{masses}) one gets a
lower bound on $M_{H_5}$ as
\beq \label{5plet} M_{H_5}\geq\sqrt{3\over 1+{t_H}^2}M_{H_3}
\; .\eeq
Constraints on $M_{H_5}$ are shown in figure(2).

"$Z^o \rightarrow {Z^o}^*\tilde{\sne },\;\tilde{\sne
}\rightarrow $ hadrons" brems\-strahlung can be used to bound
the $\tilde{\sne} $ mass from below. Results for the SM Higgs
boson \cite{DAVIER} translate to our case with help of the ratio
\beq B={\Gamma (Z^o\rightarrow {Z^o}^*\tilde{\sne })\over\Gamma
(Z^o\rightarrow {Z^o}^*H^o_{SM})}
= {c_S}^2{c_H}^2 +{8\over 3}{s_S}^2{s_H}^2 +2\sqrt{8\over
3}c_Ss_Sc_Hs_H \; .\eeq
Those bounds are displayed in figure(3). The $\Upsilon
$ decay yields the additional low energy bound \cite{RPP}
\beq M_{\tilde{\sne }} \geq\;\sim 5\; GeV \eeq
as long as the Yukawa couplings of this field are enhanced
with respect to the SM, or in other words as long as the
condition ${{c_S}^2\over {c_H}^2} \geq 1$ is met. For the case
of zero mixing this is always the case. This bound is also
valid for the field $\tilde{\snp }$ if ${{s_S}^2\over
{c_H}^2}\geq 1$.

In the limit of $t_H \ll 1$ and $c_S \approx 1$ (no
$\sne\snp $ mixing) the field $\snp $ is likely to be
lighter than the other scalar Higgs fields, but it can not
be constrained due to its strongly suppressed $HVV$ and $Hff$
type vertices. In this model, therefore, it is possible to have a
very light Higgs boson, that escapes detection.

Let's now concentrate for a moment on the properties of $\fne$.
Assuming that $M_{\fne } \leq 92 \; GeV$ weakens, as mentioned
before, the
unitarity bound on $t_H$ ($\leq 4.4$). An $\fne $ below the $Z$
threshold, therefore, could be produced by
bremsstrahlung ($Z^o\rightarrow {Z^o}^*\fne$) at LEP with
rates comparable to the SM. The
relative production rates are equal or enhanced, compared to
the SM, for $t_H$ being in the interval $[1.8,4.4]$ with a
maximal value of $1.27$ for $t_H=4.4$. Such a Higgs could also
be produced through $WW$ fusion at hadron colliders, but
there the production rate is suppressed by a factor of three or
more compared to the case of SM Higgs production.

The decay channels of $\fne $ within the Higgs sector are
$\fne\rightarrow\snp\snp $ and $\fne\rightarrow\tne\tne$ or $\tp
H_3^-$.\fn{Again we consider only the "no singlet
$\sne\snp $ mixing" sublimit.} A "close to the unitarity
limit" ($t_H > 1.8$) $\fne $ below the $W$, $\snp $ and $H_3$
threshold could therefore only decay through ${\snp}^* {\snp
}^*$, $V^*V^*$,${H_3}^* {H_3}^*$, 1-loop and $\cusym $
violating mixing diagrams, since it has no tree level
couplings to the fermion sector (eq(\ref{yukawa})). Decays to
$f\bar{f}$ pairs are mediated by triangle diagrams involving
gauge bosons and members of the $H_3$-plet and by the
$\fne\sne $ mixing term, which is taken to be of the same
order of magnitude as other 1-loop diagrams. Such a $\fne $
could therefore have a relatively large ($\sim $10\%)
branching ratio to
$2\gamma $'s\fn{Compare with the $2\gamma $ branching ratios of
the SM Higgs boson (see e.g. ref.\cite{ZOLTAN}).}, given
by 1-loop diagrams, with all the charged particles,
that couple to $\fne $, running around the loop. Again it
should be emphasized that, due to the fine tuning in this model,
the corresponding branching ratio cannot be computed.

In summary, we pointed out that in Higgs triplet models where
the $\rho\approx 1$ constraint is satisfied by imposing a
custodial $\cusym $ symmetry on the potential,
the gauge boson masses still can be dominated by
triplet contributions.
Unfortunately (or fortunately?) this idea can only be implemented
invoking fine tuning of some of the parameters of the
general $\gasym $ invariant potential, which leads to a loss
of the next-to-leading order predictability.
 The assumptions of "perturbative unification" and
"desert" then lead to a renormalization group argument which
requires that at least $\approx 20\%$ of the
$W-mass$ must come from doublet contributions.
Close to this limit  a neutral scalar with a
mass below the $W$-threshold
could have a large $B(H^o\rightarrow 2\gamma)$ branching
ratio, and could be produced with rates comparable to the SM
at LEP through $Z^o\rightarrow {Z^o}^*H^o$ bremsstrahlung. In
the other limit, with the gauge boson masses produced by the
$\sul $ doublet field, the model resembles very much to the
SM. Although here one of the neutral scalars could be very
light and still escape detection.

{\bf Acknowledgement.\ \ \ } We thank J. P. Derendinger for
illuminating
discussions and
for calling our attention to the result of paper \cite{YUKREN}.



\newpage
\noindent
{\large Figure captions}
\vskip0.8truecm
\noindent
Figures 1

\small \label{mh3} Combined bounds on
$M_{H_3}$, for $M_{H_5}$ being smaller
(A) (resp. higher (B)) than $92$ $GeV$.
"$\tan (\gt _H)\equiv t_H$" denotes the tangent of the
$\sul $ doublet triplet mixing angle as defined
in eq(\ref{ch}).
In both diagrams the high energy bound
of eq(\ref{hibound}) is represented by the dashed line, the
solid line refers to the unitarity bound of
eq(\ref{th}) and the dash-dotted line denotes the constraint
coming from $b\rightarrow s\gamma $ decay
\cite{BSY}. In both figures the shaded region is
excluded. The dotted line in picture (B)
and the upper dotted line in picture (A) represent the
$B\bar{B} $ mixing bound for "nominal" values of the
hadronic matrix elements, that enter the computation of
$B\bar{B} $ mixing, as used in \cite{BBAR1}.
For comparison the $B\bar{B} $ bound for "pessimistic" values
of the parameters (which would correspond to an actual bound) has
been plotted in figure (A) (lower dotted line). It obviously
doesn't cover any new parameter space.

\vskip 0.5truecm
\noindent
Figure 2

\small \label{mh5} Combined bounds on
$M_{H_5}$. Again the horizontal dashed line denotes the high
energy bound of eq(\ref{hibound}), whereas the
vertical solid line marks the unitarity limit on
$\tan (\gt _H)(\equiv t_H)$. The
dotted line represents the bound coming from
eq(\ref{5plet}) and eq(\ref{hibound}). The
$b\rightarrow s\gamma $ constraint as plotted in
fig(1) yields together with
eq(\ref{5plet}) the dash-dotted line. The shaded region is
excluded. Again the strict
bound from $B\bar{B}$ mixing ("pessimistic" values) covers no
new parameter space. For $\tan (\gt _H) = 0$ the
limit reads $M_{H_5} > 72\; GeV$.

\vskip 0.5truecm
\noindent
Figure 3

\small \label{mh1} Bremsstrahlung bound on
$M_{\tilde{H_1^o}}$ for the case of zero singlet mixing ($s_S
= 0$ (dashed)) and maximal singlet mixing ($s_S =
{1\over\sqrt{2} } $ (dotted), $s_S =
-{1\over\sqrt{2}}$ (dot-dashed)) and for $\tan (\gt _H)
< 2.5$, which corresponds to the stronger bound on $\tan (\gt _H)$ as
given in eq(\ref{th}). This experimental bound
represents the combined results of the four LEP experiments
for the process $Z\rightarrow Z^*H^o$ with $H^o$ decaying to
hadrons \cite{DAVIER} . In each case the region below
the corresponding line is excluded.

\end{document}